\documentclass[manuscript,nonacm]{acmart}

\usepackage{array}
\setcopyright{none}


\title{Making Generative AI Workable at Work: Accountability, Localization, and Representation in Kenyan and Nigerian Small Businesses}

\author{Najeeb Gambo Abdulhamid}
\email{nabdulhamid@microsoft.com}
\affiliation{%
  \institution{Microsoft Research Africa}
  \city{Nairobi}
  \country{Kenya}}

\author{Stephanie Nyairo}
\email{snyairo@microsoft.com}
\affiliation{%
  \institution{Microsoft Research Africa}
  \city{Nairobi}
  \country{Kenya}}

\author{Liz Ankrah}
\email{t-ankrahliz@microsoft.com}
\affiliation{%
  \institution{Microsoft Research Africa}
  \city{Nairobi}
  \country{Kenya}}

\author{Jacki O'Neill}
\email{Jacki.ONeill@microsoft.com}
\affiliation{%
  \institution{Microsoft Research Africa}
  \city{Nairobi}
  \country{Kenya}}

\hypersetup{
  pdfauthor={Najeeb Gambo Abdulhamid, Stephanie Nyairo, Liz Ankrah, and Jacki O'Neill},
  pdfkeywords={generative AI at work, accountability labour, release work, context misfit, small and medium-sized businesses, Kenya and Nigeria, human-AI interaction, localization}
}

\begin{document}

\begin{abstract}
Generative AI is widely framed as democratizing skilled work by lowering the barrier to producing professional output. Drawing on 21 semi-structured interviews with 27 workers across 21 small and medium-sized businesses (SMBs) in Kenya and Nigeria (July to October 2023), we find that availability alone did not make participation workable. In participants' accounts, drafting was rarely the hard part. The consequential work was making AI-assisted drafts defensible enough to circulate to clients. We identify five intertwined practices through which participants made generative AI workable: using it as draft labour, steering outputs through iterative interaction, switching and comparing across tools, performing release work to make outputs accountable before circulation, and repairing context misfit when outputs failed local requirements. We make two central contributions. First, we specify release work as a distinct, boundary-facing form of accountability labour, evidenced through verification gates, traceability reconstruction when sources fail, and non-delegation boundaries that decide what can ship. Second, we show how localization and representational failures, which we call context misfit, operate in everyday work as a recurring breakdown-and-repair class, including failures of local language, institutional knowledge, and the credible representation of African people and places, and we document the repair repertoire they demanded. Reframing democratization as workable participation, we argue that reported gains were conditional on this accountability and repair labour, and we translate the findings into four design directions for workplace generative AI. We treat Kenya and Nigeria as two situated cases, not as a representative sample of SMBs or the Global South.
\end{abstract}


\maketitle

\section{Introduction}

Generative AI tools are beginning to be incorporated into everyday work, often through informal, worker-led experimentation rather than top-down deployment \cite{brynjolfsson2023,wagman2025,woodruff2024}. Employees use these systems to draft, translate, design, summarize, and ideate, then integrate outputs into workflows that carry reputational, legal, and commercial consequences \cite{brynjolfsson2023,wagman2025}. For CSCW, the central question is not only what these tools can generate, but how workers make them usable, accountable, and safe within collaborative work practices \cite{orlikowski2000,star1996,suchman1987}.

Much public and organizational discussion frames generative AI as a productivity accelerant and, increasingly, as a democratizing force because it lowers barriers to experimenting with advanced capabilities through natural language interfaces and low entry costs \cite{brynjolfsson2023,seger2023,woodruff2024}. We treat democratization in a specific sense: democratization of use, meaning whether more people and more organizations can effectively apply AI in their work without specialized infrastructure or expertise \cite{seger2023}. In this paper, we treat effective use as the ability to produce locally credible, accountable outputs with acceptable time, verification effort, and risk. This matters for CSCW because access alone does not determine workplace outcomes. Technologies become workable through situated practices that reveal gaps between intended and actual use, and that redistribute responsibility for quality, verification, and risk \cite{orlikowski2000,star1996,suchman1987}. We use workable participation as an analytic lens in this study, not as a general metric of democratization across all workplaces.

In our participants' accounts of early workplace use, a recurring consequential gap was not only inaccuracy, but context misfit. In these accounts, mismatches are consistent with concerns in prior work about how training data and evaluation regimes unevenly represent regions, languages, and cultural contexts \cite{bender2021,blodgett2020,bommasani2021,joshi2020,manvi2024}. We define context misfit as recurring output mismatches along three dimensions: (1) local institutions and place, including policies, regulations, and market norms; (2) language practices, including multilingual and code-mixed communication; and (3) representational fidelity in generated language and media. In our findings, these dimensions surface as locality gaps in institutional knowledge (Section 4.5.3), language and accent breakdowns (Section 4.5.2), and representational failures in image and voice generation (Section 4.5.4).

These mismatches can produce rework, increase reputational risk, and heighten caution about what can be delegated to AI in client-facing settings \cite{brynjolfsson2023,wagman2025}. Prior workplace studies have documented how generative AI supports drafting and knowledge work, but we show how participants described a recurring coordination burden through verification, localization, and repair when outputs did not fit local institutional, linguistic, or representational expectations \cite{blodgett2020,brynjolfsson2023,joshi2020,manvi2024,wagman2025,woodruff2024}. We examine how context misfit becomes visible through everyday workplace practices of verification, localization, and repair, and what this implies for claims about democratization in practice \cite{orlikowski2000,seger2023,star1996,suchman1987}. 

We study these dynamics in small and medium-sized businesses (SMBs) in Kenya and Nigeria. In our study sites, adoption commonly occurs through pragmatic experimentation rather than formal organizational programs. SMB settings make integration work particularly visible because the same people who use the tools often also carry responsibility for client-facing quality, reputational risk, and delivery timelines. This creates clear incentives for verification, localization, and tool switching when outputs do not fit local audiences or institutional contexts. We do not treat Kenya and Nigeria as representative of the Global South. Instead, we use these sites to surface dynamics of adoption, breakdown, and repair that can be less visible when tools already align with the dominant language, institutional sources, and representational defaults of a given setting \cite{blodgett2020,joshi2020,manvi2024}. 

We report findings from a qualitative interview study of generative AI use in SMB workplaces in Kenya and Nigeria. We conducted 21 semi-structured interviews with 27 participants from 21 SMBs between July and October 2023. Participants described using multiple tools, primarily for text and occasionally for image and speech generation and provided concrete accounts of where these tools created value and where they failed. Our analysis focuses on the practical work of integration: how outputs are verified, localized, and stabilized for workplace use.

We ask:

RQ1. How do SMB workers in Kenya and Nigeria incorporate generative AI tools into everyday work, and what practices emerge around tool selection, verification, and accountability before client-facing release?

RQ2. Where does context misfit surface in generative AI use, and what strategies do workers use to repair and localize outputs to meet workplace standards?

Our analysis surfaces five often intertwined practices that structured generative AI use in SMBs in Kenya and Nigeria. These include: treating generative AI as draft labour for client-facing deliverables; steering output quality through interaction routines (including rewriting, prompt adjustments, and voice-preserving edits); distributing use across an ecology of tools through switching and comparison; performing release work to make outputs accountable before circulation (including verification gates, traceability reconstruction, and non-delegation boundaries); and repairing context misfit when outputs failed to match local institutions, language practices, or representational expectations. Together, these practices complicate simple access-and-productivity narratives by showing that tool availability did not by itself make participation workable because outputs still had to be made defensible for circulation.

Contributions. This paper makes two central contributions to CSCW research on generative AI at work, connected by a reframing of democratization and supported by a practice account and design directions.

Our first contribution is conceptual. We specify release work as a distinct, boundary-facing form of accountability labour that determines whether an AI-assisted draft can circulate to an external audience. In our data it is evidenced through verification gates tied to domain standards, traceability reconstruction when references fail, and non-delegation boundaries that allocate responsibility for what ships. Where articulation work names the ongoing coordination that meshes interdependent tasks and actors so that cooperative work coheres, release work names something narrower: the labour of certifying a single artifact as defensible before it crosses to an external audience. Release work can be performed by one worker without coordinating others, and is triggered by external accountability at the point of circulation rather than by task interdependence, so it is not reducible to articulation work.

Our second contribution is empirical. We show how localization and representational failures, which we call context misfit, operate in situated workplace use as a recurring breakdown-and-repair class that gates circulation. This contribution is mechanistic rather than taxonomic: we do not claim a new category of bias, since prior work already documents geographic, cultural, linguistic, and representational bias. What we add is an account of how these failures play out in everyday work, where outputs failed local institutions and place knowledge, local language and accent practices, and the credible representation of African people and places. We document the repair repertoire this demanded, including rewriting for voice, constraint layering, substituting local sources and situated knowledge, tool switching, and disengagement when repair costs outweighed value, making visible the labour that decontextualized benchmark evidence does not capture.

We connect these through workable participation, an analytic reframing of democratization from access toward the ability to produce locally credible, accountable outputs within acceptable time, verification effort, and risk. Supporting these contributions, we give a five-practice account of how drafting, steering, ecological tool use, release work, and context-misfit repair were arranged as outputs moved toward circulation, and we translate the findings into four design directions for workplace generative AI.

The remainder of the paper reviews related work, describes our methods and analysis, presents findings organized around these practices and breakdown and repair dynamics, and discusses implications for CSCW research and design directions for workplace generative AI.

\section{Literature}

\subsection{Working with Generative AI}

CSCW has long shown that workplace technologies rarely enter work as neutral add-ons. Their value and consequences are produced in practice, through appropriation, articulation work, coordination mechanisms, and repair when breakdowns occur in real settings \cite{orlikowski2000,schmidt1992,schmidt1996,star1996,suchman1987}. This practice lens is especially useful for generative AI (GenAI) because these systems do not only ``complete tasks.'' They introduce a new interactional resource whose outputs are easy to reuse, but whose correctness, fit, and accountability are often uncertain, shifting what workers must coordinate, document, and justify to get work done \cite{orlikowski2000,schmidt1992,schmidt1996,star1996,suchman1987}.

Empirical evidence on GenAI-at-work reports meaningful average productivity gains in some settings, while also highlighting strong heterogeneity across workers and tasks. Field evidence from customer support finds improvements on average, with substantially larger gains for novices and smaller or negligible effects for highly experienced workers \cite{brynjolfsson2023}. Experimental studies of professional writing similarly report faster completion and quality improvements on average, while raising questions about how work is reorganized toward drafting, editing, and evaluation rather than ``from-scratch'' production \cite{noy2023}. These results motivate a CSCW focus on how GenAI changes workflow organization, including who benefits, what kinds of oversight are required, and where responsibility for checking and sign-off sits in everyday practice \cite{brynjolfsson2023,noy2023}.

A complementary line of work characterizes GenAI's impact as uneven along a ``jagged frontier,'' where outcomes depend on whether tasks can be decomposed into verifiable steps, and whether organizations have stable criteria for assessing outputs \cite{dellacqua2023}. From this view, the limiting factor is often not generation but verification. Workers must decide when an output is good enough to reuse, when to escalate, and when to switch tools or revert to other sources. These are not peripheral behaviours. They are coordination mechanisms that shape time, risk, and accountability in collaborative work \cite{dellacqua2023,star1996}.

HCI and CSCW studies of GenAI use in real-world practice further show that workers develop interactional strategies to stabilize output quality, including iterative prompting, adding constraints and examples, breaking tasks into subtasks, and triangulating across tools or sources \cite{sun2024}. These strategies frequently become social rather than purely individual, as teams share prompt templates, negotiate what counts as ``acceptable,'' and routinize checks that protect reputational and operational risk. In this sense, GenAI can reduce first-draft effort while increasing the articulation work around quality control and traceability \cite{star1996,sun2024}.

Software development illustrates this reconfiguration clearly. Studies of code assistants report perceived speed and flow benefits, but also document that generated code requires careful review and testing, and can introduce security-relevant weaknesses that are not immediately obvious from fluent output \cite{bird2022,pearce2025}. The practical consequence is a redistribution of effort: less time producing an initial draft and more time validating, debugging, and managing organizational guardrails. This aligns with classic CSCW arguments that ``getting work done'' is a collective achievement sustained through coordination and repair, not simply individual productivity \cite{pearce2025,schmidt1992,schmidt1996,star1996}.

Finally, these dynamics are inseparable from infrastructural and contextual conditions. Ethnographic work on workplace communication in India and Kenya shows how tools are embedded in layered communication ecologies shaped by device constraints, mobility, and the rhythms of local and remote work \cite{mcgregor2019}. Such conditions are likely to interact with GenAI by shaping tool choice, verification pathways, and the social organization of checking, especially in multilingual settings where workers move across languages, platforms, and locally grounded knowledge. This motivates examining GenAI in SMB workplaces as part of a broader tool ecology, and organizing analysis around mechanisms such as decision-making, verification, breakdowns, and repair. It also sets up the next subsection, where we review how representational and localization limits in GenAI become consequential through these workplace mechanisms.

\subsection{Representation and Bias in Generative AI Models}

Concerns about representation and bias in AI systems are longstanding, and they take on heightened significance for generative AI because these models produce fluent, plausible outputs that can be quickly reused in communication, decision support, and content production \cite{bender2021,gallegos2024,weidinger2021}. Systematic accounts further emphasize that ``fairness'' problems in generative AI are rarely singular, and often emerge from interacting choices across data, modeling, evaluation, and deployment \cite{elmimouni2025,weidinger2021}.

A consistent explanation in the literature is that representational bias is shaped by the data and development pipeline. Training data reflects uneven visibility of people, places, and language varieties in the digital record, and models can therefore privilege dominant perspectives while marginalizing or distorting less represented ones \cite{bender2021,weidinger2021}. These issues are not limited to overt stereotyping. They also include subtle forms of omission and ``thinness,'' where generated outputs sound confident but fail to reflect locally grounded institutions, practices, or culturally situated meanings \cite{adilazuarda2024,weidinger2021}. Work on evaluation highlights a related problem: benchmark findings depend on task design, datasets, metrics, and the contexts represented, so common testing regimes can under-detect place-based or culturally specific failures \cite{eriksson2025,ochieng2024,phutane2025}.

Language is a particularly visible axis of uneven representation. Reviews of low-resource language support show persistent gaps in data availability, tooling, and evaluation coverage, which can compound exclusions for communities whose language varieties are rarely captured in high-volume training corpora \cite{nekoto2020}. Recent benchmarks of LLMs on African languages further underscore that ``multilingual capability'' is not evenly distributed, and that performance variability can remain significant even when systems claim broad language coverage \cite{ifeoluwaadelani2024,ojo2025}. Complementary interventions argue for explicitly addressing structural and ethical issues in NLP development for low-resource settings, including the relationships between language, power, and whose knowledge becomes legible to computational systems \cite{joshi2020}. In parallel, multilingual and context-sensitive evaluation efforts emphasize that testing needs to move beyond English-centric assumptions and attend to how systems behave across a wider range of linguistic contexts and communicative norms \cite{ifeoluwaadelani2024,ochieng2024,ojo2025}. More recent work also makes visible how ``bias'' and ``representation'' concerns connect to the practical availability of African-language evaluation resources and to the kinds of social biases surfaced in language models for African settings \cite{beux2025,ifeoluwaadelani2024,ochieng2024,ojo2025}.

Representational problems also appear in generated media, where systems may reproduce narrow portrayals of people and places or re-inscribe dominant visual narratives. For example, work in global health visuals shows how text-to-image systems can reproduce biased imagery rather than disrupt it, even when users explicitly attempt to counter stereotypes \cite{alenichev2023}. Related community-centered analyses demonstrate how generative models can preserve an ``outsider's gaze'' in cultural depictions, reflecting broader power asymmetries in whose perspectives are encoded and reproduced \cite{qadri2023}. These representational concerns are compounded by intersectional dynamics, where region, race, gender, and class can interact in ways that intensify harms or exclusion \cite{crenshaw2013}.

Across audits, benchmark suites, and survey work on bias, a recurring critique is that decontextualized evaluation can miss how representational and localization gaps surface in everyday use, especially in culturally nuanced and low-resource settings \cite{adilazuarda2024,beux2025,eriksson2025,ochieng2024,phutane2025}. For generative systems, this matters because breakdowns are often partial rather than clearly ``wrong.'' Outputs can read as fluent and plausible while remaining thin on locally salient entities, defaulting to globally dominant framings, or failing to retrieve long-tail, place-specific knowledge that users implicitly rely on \cite{kandpal2023,ochieng2024}. These partial mismatches are also more likely when users work through language varieties, dialects, and code-mixed expression that are unevenly represented in training data and evaluation regimes \cite{ifeoluwaadelani2024,nekoto2020,ojo2025,zhou2025}. In work settings, these partial mismatches become consequential when they require verification, repair, and coordination to make outputs defensible for release and circulation.

Building on calls to better conceptualize and measure ``culture'' in LLM research without reducing it to simplistic proxies \cite{adilazuarda2024}, this paper uses context misfit in a deliberately narrow sense: recurring localization and representation failures that become visible through interaction with generative AI, rather than an essentialized claim about any region. In this section we focus on localization and representational misfit, which prior audit and benchmark work sometimes discusses in terms of geographic or regional bias. We operationalize this form of context misfit along dimensions that can be observed in use, including: (a) under-localized place knowledge and named entities, including long-tail and locally salient information \cite{kandpal2023}; (b) culturally misaligned or externally shaped representations of people and places in generated content \cite{alenichev2023,basu2023,zhang2024}; (c) uneven support for language varieties and code-mixed expression, including lower-resource languages and dialectal variation \cite{ifeoluwaadelani2024,joshi2020,nekoto2020,ojo2025,zhou2025}; and (d) limitations in accent and dialect handling where speech interfaces are involved \cite{koenecke2020,tatman2017}. We examine how these misfits are encountered and worked around in SMB work practices in Kenya and Nigeria.

\subsection{AI divides as workplace constraints}

Digital divide research frames inequality as more than connectivity, emphasizing multifaceted disparities that shape who can translate digital technologies into meaningful outcomes \cite{vassilakopoulou2023}. Work on AI and the digital divide similarly stresses that participation in the ``AI revolution'' depends on enabling conditions rather than simple availability \cite{bentley2024,kitsara2022}. Infrastructure scholarship further shows that technologies are embedded in wider arrangements and shaped by non-local constraints, so making systems workable often requires negotiating these constraints in practice \cite{monteiro2013,star1996}.

A growing body of work argues that generative AI may replicate and intensify existing inequalities rather than automatically democratize advanced capabilities \cite{capraro2024,daepp2024}. At a structural level, divides reflect uneven foundations such as access to resources, institutional capacity, and the sociotechnical conditions required to adopt, govern, and benefit from AI systems. These dynamics matter for workplaces because they can translate into uneven reliability, uneven perceived value, and uneven ability to realize productivity gains, even when awareness and exposure are widespread \cite{gmyrek2024}.

For SMBs, we treat constraints as decision conditions that shape how generative AI can be incorporated into everyday work. Pricing and feature gating can make effective use contingent on free tiers, trials, or limited tool capabilities, which can shape tool choice and encourage cross-tool workflows. Privacy, confidentiality, and regulatory uncertainty can also shape what kinds of prompts workers consider acceptable, when they avoid sharing sensitive materials, and what forms of obfuscation or redaction become part of routine use \cite{hacker2023}. Finally, CSCW research on workplace communication in the Global South highlights how work is often organized through layered tool ecologies and cross-platform coordination, which can become the practical substrate through which new tools are tried, combined, and managed \cite{mcgregor2019}.

Taken together, this lens shifts the analytic question from whether generative AI is ``available'' to how workers navigate constraints through tool selection, disclosure choices, and verification routines under time pressure and reputational risk. In this paper, we use ``AI divides'' in a deliberately practice-oriented way: to explain how cost, privacy, infrastructure, and capability constraints become visible as (1) tool selection and switching strategies, (2) selective disclosure and data-handling decisions, and (3) verification burdens and trust calibration as ongoing workplace work. These are the mechanisms we return to when interpreting how SMB workers make generative AI usable and accountable in practice.

\section{Methodology}

\subsection{Research Sites and Scope}

This study focuses on workplace use of generative AI among SMB workers in Kenya and Nigeria. We selected these sites for analytic leverage rather than representativeness. At the time of data collection (2023), Kenya and Nigeria offered a useful comparative lens because both are established digital innovation hubs with active startup ecosystems and substantial SMB activity, creating plausible conditions for early workplace experimentation with new AI tools \cite{startupblink2023}. Both countries also score meaningfully on widely used ``AI readiness'' indicators, suggesting enabling conditions for AI experimentation and uptake \cite{hankins2023}. More broadly, their mobile-first connectivity environments support day-to-day digital work practices through which generative AI tools can be accessed and integrated \cite{gsmaintelligence2023}. 

Again, both are multilingual workplace ecologies where English is commonly used for business, while local languages, place-based knowledge, and culturally specific references remain embedded in everyday work. This allowed us to observe routine uses of generative AI as well as context-sensitive breakdowns, including failures around local language translation, local markets and places, culturally specific narratives, identity depiction, and accent support. We do not claim that Kenya and Nigeria represent Africa or ``the Global South.'' Instead, we treat them as two situated cases that illuminate how generative AI is integrated into SMB work and how context-misfit can shape that integration. 

We also selected these sites for practical access reasons: as some members of the research team are originally from Kenya and Nigeria, providing familiarity with local contexts and access to professional networks that facilitated participant recruitment.

\subsection{Defining SMBs and Selecting Organizations}

We use the term ``small and medium-sized business (SMB)'' as an operational category for the organizations represented in our sample. In this study, an organization qualified as an SMB if it employed fewer than 250 people, aligning with the European Commission's definition \cite{europeancommission2003}. We also considered African Union guidance on MSME definitions, which notes that MSME thresholds vary across contexts and provides an adopted framing that includes turnover-based categories and commonly used employee thresholds \cite{africanunioncommission2022}. We adopt the <250 threshold as a pragmatic criterion to support comparability across diverse sectors in our sample; our goal is operational consistency rather than alignment with any single national standard. We determined SMB status through a combination of self-reported organizational size and participants' descriptions of their roles and organizational structure during recruitment. In many cases, participants explicitly identified their organization as a small or medium-sized enterprise, and their accounts of team size, business structure, and operational scope were consistent with this classification. 

We focus on SMBs because participants described generative AI use as tightly coupled to everyday delivery constraints, including limited time, client-facing accountability, and selective disclosure of sensitive information. This makes the work of integrating, validating, and repairing AI output visible in routine practice. 

\subsection{Participants and Organizations}

We conducted 21 semi-structured interviews with 27 participants drawn from 21 SMBs in Kenya and Nigeria between July and October 2023. Twenty interviews were conducted one-to-one, and one was a group interview with seven participants from the same legal-services firm (RP17 and RP17A--RP17F). Participants were distributed across the two sites as follows: Kenya (n = 18) and Nigeria (n = 9). The 21 SMBs spanned multiple sectors (Table 1), including professional services (n = 7), creative/media (n = 6), retail/commerce (n = 3), technology (n = 3), training (n = 1), and other services (n = 1). 

To make heterogeneity analytically legible, we report role buckets that we use explicitly in the analysis. By organizational position, our sample included owners/founders (n = 12) and employees/managers (n = 15). By work orientation, most participants held client-facing roles (n = 21), while a smaller subset held back-office roles (n = 6). Table 1 reports participant roles, sectors, and generative AI tools used. 

Sampling note. Because eligibility required recent workplace GenAI use, this sample reflects SMB professionals already using these tools and should not be read as prevalence or adoption estimates.

\begin{table*}[t]
\centering
\caption{Participants and the generative AI tools they use.}
\label{tab:participants}
\footnotesize
\setlength{\tabcolsep}{3pt}
\begin{tabular}{>{\raggedright\arraybackslash}p{0.080\linewidth}>{\raggedright\arraybackslash}p{0.150\linewidth}>{\raggedright\arraybackslash}p{0.150\linewidth}>{\raggedright\arraybackslash}p{0.070\linewidth}>{\raggedright\arraybackslash}p{0.220\linewidth}>{\raggedright\arraybackslash}p{0.120\linewidth}>{\raggedright\arraybackslash}p{0.110\linewidth}}
\toprule
\textbf{Participant ID} & \textbf{Job sector} & \textbf{Title} & \textbf{Country} & \textbf{Generative AI tools used} & \textbf{Organizational position} & \textbf{Work orientation} \\
\midrule
RP13 & Architectural Design & Product Designer & Kenya & Leonardo.Ai, Adobe Firefly, ChatGPT, Midjourney, BrieflyAI & Employee & back-office \\
RP14 & Outdoor recreation & Guide & Kenya & Google Bard, ChatGPT & Employee & client-facing \\
RP15 & Retail & CEO & Kenya & Runway, Canva, Descript, BrowseAI, Naming Magic, ChatGPT & Owner & client-facing \\
RP16 & Retail & Managing Director & Kenya & ChatGPT & Owner & client-facing \\
RP17 & Legal Service & Partner & Kenya & ChatGPT, Bing Chat, Midjourney & Employee & client-facing \\
RP17A & Legal Service & Head of Business Development & Kenya & ChatGPT, Bing Chat, Midjourney & Employee & client-facing \\
RP17B & Legal Service & Trainee Associate & Kenya & ChatGPT, Bing Chat, Midjourney & Employee & client-facing \\
RP17C & Legal Service & Associate & Kenya & ChatGPT, Bing Chat, Midjourney & Employee & client-facing \\
RP17D & Legal Service & Intern & Kenya & ChatGPT, Bing Chat, Midjourney & Employee & client-facing \\
RP17E & Legal Service & Trainee Advocate & Kenya & ChatGPT, Bing Chat, Midjourney & Employee & client-facing \\
RP17F & Legal Service & Associate & Kenya & ChatGPT, Bing Chat, Midjourney & Employee & client-facing \\
RP18 & Printing \& Design & Research and Development Lead & Kenya & Photoshop Beta, Adobe Firefly, Midjourney, ChatGPT & Employee & client-facing \\
RP19 & Creative Writing & CEO/\allowbreak{}Founder & Kenya & ChatGPT, Canva & Owner & back-office \\
RP20 & Outdoor Recreation & CEO & Kenya & ChatGPT, Canva & Owner & client-facing \\
RP21 & Real estate & Founder/\allowbreak{}Director & Kenya & ChatGPT & Owner & client-facing \\
RP22 & Imports and distribution of Goods & Sales and Marketing Lead & Kenya & ChatGPT, DALL-E & Employee & client-facing \\
RP23 & Financial Services & Operations Manager & Kenya & ChatGPT & Employee & back-office \\
RP24 & IT & CEO/\allowbreak{}Founder & Kenya & ChatGPT, Leonardo.Ai, Stable Diffusion, ElevenLabs, Midjourney & Owner & client-facing \\
RP25 & IT & Software Engineer & Nigeria & ChatGPT, Bing, Synthesia, Pictory & Employee & client-facing \\
RP26 & Clothing \& Fashion Design & CEO & Nigeria & ChatGPT & Owner & client-facing \\
RP27 & Business Development & Founder & Nigeria & ChatGPT & Owner & client-facing \\
RP28 & Digital Marketing & CEO & Nigeria & ChatGPT, Midjourney, Google Bard & Owner & back-office \\
RP29 & Financial Services & Lead Consultant & Nigeria & ChatGPT & Owner & back-office \\
RP30 & General contract \& supplies & Partner & Nigeria & ChatGPT, Google Bard, Bing Chat & Owner & client-facing \\
RP31 & Multimedia Journalist & COO & Nigeria & ChatGPT & Employee & client-facing \\
RP32 & Power Sector & Lead Trainer & Nigeria & ChatGPT, Google Bard, Bing Chat & Employee & client-facing \\
RP33 & Telecommunication & CEO & Nigeria & ChatGPT, Leonardo.Ai, Midjourney, Blue Willow & Owner & back-office \\
\bottomrule
\end{tabular}
\end{table*}

\subsection{Recruitment}

Participants were recruited through LinkedIn, WhatsApp groups, professional associations, and snowball referrals. Eligibility required that participants (a) currently worked in an SMB as defined above and (b) had used at least one generative AI tool for work tasks at least three times per week over the past six months. Brief screening questions confirmed eligibility and captured basic context (role, sector, tools used). Because fieldwork ran from July to October 2023, within the first year of widespread public access to consumer generative AI (ChatGPT launched November 2022), participants were regular but early-stage users. We did not record exact length of use, which we note as a limitation.

This approach surfaces integration, breakdown, and repair mechanisms that can be less visible in settings where dominant language and cultural assumptions align more closely with mainstream tools. 

\subsection{Data Collection}

Interviews lasted 60--90 minutes and were conducted in English. This may privilege participants with higher comfort using English at work and may under-observe prompting and collaboration practices that occur primarily in local languages or through code-mixed interaction. We therefore interpret language-related practices and breakdowns as those participants were able to report and demonstrate within English-language interviewing, rather than as exhaustive accounts of multilingual workplace interaction. Three interviews were conducted remotely via Microsoft Teams; the remaining 18 interviews took place in-person at participants' workplaces or at the researchers' office, depending on participant preference and accessibility. Interviews followed a semi-structured protocol covering four themes: AI discovery and tools, AI impacts, AI skills, and AI vision, with follow-up probes on verification practices, data sensitivity, and breakdowns encountered in locally grounded work.

Participants were invited to demonstrate how they used generative AI tools in their work, including showing prompt histories, outputs, and interaction flows when available. These demonstrations varied by participant comfort, workplace setting, and device access, and we treat them as supportive context rather than a uniform observational record. We did not standardize demonstrations across participants, and we rely on interview accounts as the primary evidence base. We did not systematically quantify modality use; references to text, image, and speech reflect what participants chose to discuss in interviews and the examples they volunteered. Accordingly, the findings use modality examples illustratively to explain practices, not to estimate prevalence. With explicit verbal consent, 9 participants shared artifacts, including screenshots or image outputs, and in one case a participant provided example outputs directly. Artifacts were used to support interpretation during analysis sessions (e.g., understanding interaction flows and output characteristics) and were not treated as a standalone dataset. Artifacts were stored temporarily in our research SharePoint for the duration of analysis and were deleted immediately after analysis was completed, in accordance with our data retention protocol. All identifying details were masked to protect participant confidentiality prior to analysis.

\subsection{Transcription and Qualitative Analysis}

Interviews were transcribed using HeyMarvin. Each transcript was checked in full against the corresponding audio recording before coding. We used HeyMarvin for transcription and organization (e.g., tagging and retrieving excerpts), not for generating codes, themes, or interpretations. We analyzed the interview data using codebook thematic analysis, a team-based approach in which a shared, evolving codebook of explicitly defined codes supports consistent application across multiple coders \cite{braun2006,guest2011,macqueen1998}. Consistent with this approach, we combined an inductive pass over the dataset with subsequent structured coding aligned to our research questions. Four authors contributed to the broader study: the first and second authors conducted the interviews, with the fourth author joining a limited number; the fourth author did not participate in the coding process. The third author participated in all stages of the research except conducting interviews and contributed to the thematic analysis by tagging a subset of the transcripts. Across analysis, we held recurring group sessions in which the first author presented candidate observations from a focal thematic area and all participating analysts jointly interrogated, cross-checked, and compared interpretations before agreeing on analytic decisions.

Phase 1: inductive coding and theme development. To maintain contextual integrity and avoid premature narrowing, the first and second authors independently reviewed and coded the interview transcripts to identify recurring patterns, practices, and breakdowns in how participants used generative AI at work. The analysis team then met to discuss overlaps and divergences and to align on a preliminary set of themes and subthemes. We maintained an evolving codebook in a shared spreadsheet accessible to all authors. The spreadsheet included code names, short definitions, and illustrative quotes to support consistent application and to document how codes evolved over time. 

Phase 2: structured coding and synthesis. After establishing the initial thematic landscape, we conducted focused coding across the dataset using an excerpt-based workflow. We treated excerpts as eligible for focused coding when they directly addressed the research questions or spoke to the core analytic themes identified in Phase 1 (e.g., iterative prompting and post-editing, cross-tool checking and tool selection, verification and trust calibration, selective disclosure of sensitive data, and context-specific failures around local language, place knowledge, identity depiction, and accent support). Excerpts were selected from across the full transcript set using shared inclusion criteria recorded in the codebook, and we revisited full transcripts during refinement to confirm that themes fit beyond selected excerpts. To reduce selection effects in excerpting, we used the codebook to retrieve additional instances of each practice across transcripts and checked whether themes held beyond the most vivid examples. The first and second authors, with support from the third author (who tagged a subset), systematically tagged these excerpts using HeyMarvin and organized them under shared codebook categories, supporting traceability from findings back to source transcripts.

Iterative refinement and disagreement handling. We conducted two rounds of theme refinement after the initial team alignment. In each round, we revisited tagged excerpts, refined code definitions, and adjusted thematic boundaries based on recurring patterns and analytic discussion. Disagreements about excerpt categorization or interpretation were surfaced during group analysis sessions and resolved through consensus, with changes recorded in the shared codebook to preserve consistency. We relied on negotiated consensus and definitional clarity in the shared codebook, rather than an inter-rater reliability coefficient, to reconcile coding decisions, consistent with reporting norms for qualitative research in CSCW and HCI \cite{mcdonald2019}.

Credibility checks. We actively searched for negative and deviant cases, defined as participant accounts that contradicted or diverged from dominant patterns. These cases were retained and discussed in team sessions and used to refine theme boundaries, surface exceptions, and avoid overgeneralization. We also conducted cross-role comparisons (e.g., business owners vs employees; client-facing vs back-office roles) to examine how practices and breakdowns varied by work position. We treated sufficient thematic density as reached for the themes reported in this paper when the second refinement round no longer yielded substantively new themes or subthemes within our dataset, and we used deviant-case analysis to probe boundary conditions and refine interpretations. Given variation across sectors and the uneven distribution across sites, we treat our results as practice-oriented accounts of practice rather than an exhaustive map of all SMB use patterns. Accordingly, we do not treat the analysis as evidence for cross-site differences.

\subsection{Ethics, Consent, and Compensation}

The study was reviewed and approved by the Microsoft Research Ethics Review Program (IRB). Participation was voluntary. We obtained informed written and verbal consent prior to each interview and reminded participants they could decline to answer any question or stop the interview at any time. We offered shopping vouchers as a token of appreciation. Because interviews included workplace practices and, at times, sensitive information, we encouraged participants to avoid sharing confidential client data and focused discussion on processes, decisions, and de-identified examples where possible. 

\section{Findings}

We report five practices through which participants described making generative AI workable for client-facing work: (1) drafting with AI to produce a first pass, (2) steering outputs through iterative interaction routines, (3) switching across tools to compare and arbitrate quality, (4) performing release work to make outputs safe to ship before circulation, and (5) repairing context misfit when outputs exhibited localization or representational failures that required repair before external circulation, or when repair costs led to abandonment.

Our analytic unit is not ``use'' in the abstract but circulation. Participants described AI drafts becoming valuable primarily when workers could assemble a safe-to-ship output package for client-facing release, shifting effort from producing text to producing accountability through verification gates, traceability reconstruction, and non-delegation boundaries. When outputs failed local fit, participants described repair through localization or substitution, and in some cases abandonment.

\subsection{Generative AI as draft labour for client-facing deliverables}

Many participants described using generative AI less as a standalone ``creative tool'' and more as a way to produce a fast first draft of materials they needed to share with clients, customers, or public audiences. The value of generative AI in these settings was not simply producing content but producing a workable starting point that participants could then localize, edit, and align with professional voice and audience expectations. In these accounts, draft labour appears as (1) drafting under delivery pressure, (2) producing structured materials for later tailoring, (3) tone calibration to make messages ``safe to send,'' and (4) a boundary case where limited expertise makes evaluation and verification particularly salient.

\subsubsection{Drafting content under delivery pressure}

Participants often used text generation to draft or unblock client-facing writing when time or phrasing was a constraint. For example, one legal advocate described using the system when stuck: ``if you have creative block and you just can't think of that clause\ldots{} I use it as just to trigger my mind and then I'll probably rephrase the entire clause.'' Here, generative AI functioned as a starting point that participants treated as material to be rewritten and held accountable to professional standards, rather than as a final product.

Similarly, participants used AI to generate starting points for talks, pitches, and outreach, then grounded them in local examples. A business development CEO who also hosts a weekly radio segment for entrepreneurs described using ChatGPT to generate ``some simple topics that will resonate with my listeners,'' then searching for real case studies from the Hausa-speaking community to illustrate them, noting that previously it was difficult to get ideas that aligned with local entrepreneurs. Participants frequently framed this workflow as a way to reach a usable draft faster. A financial services consultant, for instance, said it ``increases the speed of my work\ldots{} things that would normally take two hours\ldots{} I can do in thirty minutes\ldots{} or maximum forty minutes,'' highlighting a perceived payoff of faster turnaround, while still requiring revision and localization to fit audience expectations and local context. These perceived speed gains were not uniform across contexts. As we discuss below, participants sometimes limited what they delegated to AI when confidentiality mattered and cautioned that drafts are harder to evaluate when the topic falls outside their expertise.

\subsubsection{Producing structured artifacts, then tailoring for fit}

This draft labour extended beyond drafting messages to generating structured materials that participants could adapt into workplace artifacts. A power sector lead trainer described using generative AI to prepare examination materials by outlining key topic areas and then asking the system to generate ``the twenty-five questions'' and their answers across those areas. Here, the key move was rapid generation of a question set that the trainer could then review, adjust, and reuse for assessment delivery.

Several participants described these workflows as faster than prior approaches, but their comparisons were typically about reaching a usable set of options or an initial output, not about replacing end-to-end work. For instance, an architect-designer said they previously had to ``bounce off ideas'' across multiple sessions, whereas now they could arrive at a similar set of candidate ideas in one session. Even in these accounts, the generated output still required human selection and refinement to fit the specific project and audience. As one participant put it, they ``already know sort of what we are looking for,'' using AI to accelerate getting there rather than to substitute for professional judgment.

\subsubsection{Polishing communication to make it ``safe to send''}

A common use of generative AI in participants' accounts was not simply ``writing emails,'' but using AI to make client communication safer to send when tone, clarity, or language carried relationship risk. Participants described email as a site where small wording choices could escalate misunderstandings, signal disrespect, or weaken credibility, especially when requesting payment, setting boundaries, or delivering bad news. Here, the key move was tone calibration: generating candidate phrasings that could be edited into an acceptable message.

Participants used the tools to condense wordy messages, correct English, and adopt a professional tone with less effort, while still revising outputs to preserve personal voice. A sales lead who is dyslexic explained, ``I use a lot of AI or ChatGPT to answer most of my emails\ldots{} it really gives me an edge of being very professional using very, very good English,'' but also noted the need to correct the output to ``sound like me.'' This highlights a workflow in which generative AI provides linguistic polish as a starting point, while participants retain accountability for voice, intent, and appropriateness.

This practice became particularly salient in delicate client situations where participants wanted to be firm without sounding rude. A legal services advocate described using the system to negotiate this boundary: ``how do I tell my client he's four months late now to this to pay his fees and I don't want to sound rude?'' Rather than outsourcing the interpersonal work, participants described using AI to generate candidate phrasings that they could adjust, allowing them to move forward without spending extensive time composing or second-guessing tone.

Participants also framed this ``safe-to-send'' workflow as a route to faster turnaround that could shape how clients perceived them. A power sector trainer described how accelerated drafting affected responsiveness: ``Somebody will request for something from you, expecting it in three days, and they get it in three hours\ldots{} they see you as someone who is actually prompt with good time management.'' In this sense, the benefit was not only time saved, but the ability to meet professional expectations for reliability and service, while still doing the human work of reviewing and tailoring messages before sending. However, this reliance on AI for outward-facing communication also heightened participants' attention to checking and revising outputs, since mistakes or misfit phrasing would be visible to clients and could damage trust.

This delegation was selective, however. A sales and marketing lead in an imports and distribution business explained that while they could ask ChatGPT to ``draft an email,'' they would not provide sensitive business information, noting, ``give you all my monthly sales\ldots{} I don't do that with ChatGPT\ldots{} our critical data, we don't give it.'' This boundary case shows how ``safe-to-send'' practices can coexist with selective withholding of confidential data, shaping what kinds of drafting work participants were willing to delegate.

\subsubsection{Boundary case: using generative AI where expertise is thin}

In a smaller set of accounts, participants described using generative AI not only to accelerate drafting, but to compensate for limited domain expertise. A contract and supplies partner, reflecting on a joint proposal written in an area where they had ``no deep experience or understanding,'' argued that the use of AI tools enabled them to contribute ``far far beyond what many other people that have even the experience in the area were able to contribute.'' This account suggests that participants sometimes treated generative AI as a scaffold for participation in unfamiliar domains, allowing them to contribute in ways they framed as exceeding what would otherwise be possible despite limited prior understanding.

At the same time, this case is analytically ambiguous from interview data alone. While the participant framed the outcome as successful contribution, the account also surfaces a boundary condition for ``draft labour'' practices: when domain expertise is thin, it may be harder to judge whether an output is merely well-phrased versus substantively correct or appropriate. A power sector lead trainer articulated this limit directly, warning that if AI is used in areas ``that are not your area of expertise\ldots{} you cannot review it, you cannot check,'' and that this increases the risk of ``false information,'' recommending that people use it within their area of expertise. This ambiguity also helps explain why participants emphasized verification, particularly when working outside their core expertise. A telecom CEO cautioned that ``The content may not be a hundred percent accurate\ldots{} you have to check that content again,'' while a financial services lead consultant described a routine of validation: ``I always cross check whenever I got response\ldots{} make sure that it is correct before I use it.'' We return to this evaluability problem in our account of verification, where participants describe how they check, revise, and validate AI-generated material before using it in accountable work, particularly when outputs could be consequential or publicly visible.

\subsection{Interaction routines for iteratively steering output quality}

Participants described that getting usable value from generative AI was rarely a one-shot prompt. Instead, they developed interaction routines to steer outputs toward what they considered usable, credible, and appropriately voiced for workplace contexts. Across these accounts, ``quality'' was not only about fluency, but about whether an output could be trusted, fit-for-purpose, and safe to circulate in workplace settings. Participants improved quality through meta-level prompting to learn how to ask (4.2.1), scaffolding prompts with templates or plugins (4.2.2), iterating through successive prompting and revision (4.2.3), and editing drafts to preserve originality and voice (4.2.4). At the same time, participants also described boundary conditions where steering did not resolve uncertainty and they reverted to manual methods, avoided using AI for particular domains, or treated outputs as unaccountable until verified (4.2.5).

\subsubsection{Learning how to ask: meta-prompting as a practical skill}

For some participants, the first challenge was not writing the content, but figuring out how to ask for it in a way that produced usable output. A power sector lead trainer described directly asking the system for guidance on prompting: ``What is the best way to ask ChatGPT\ldots{} so that we can get a high-quality output?'' Similarly, another participant described turning to AI when they did not know how to phrase a complex request, asking: ``With these complexities, what is the best way of creating a prompt? Can you generate a prompt?'' In these cases, prompting becomes part of the work practice itself, where participants use the system to learn how to elicit better outputs for later tasks.

\subsubsection{Scaffolding prompts with external tools and templates}

Participants also described using external scaffolds to push outputs toward workplace-acceptable style, especially when default responses felt formulaic or visibly ``AI-like.'' A telecom vendor emphasized the use of plugins such as AIPRM to shape outputs into what they called ``human like content,'' explaining: ``Honestly most of them are human like content\ldots{} Hundred percent human like content that's why we emphasise on the use of AI PRM.'' Here, ``quality'' is framed as plausibility for audiences and reputational safety, not only grammatical correctness.

\subsubsection{Iterative refinement loops: prompting as editorial work}

Participants frequently described output quality as something produced through successive adjustments rather than retrieved in a single response. One digital marketer described this iterative loop plainly: ``If I don't like [the response], then I add another prompt to modify the previous response until the response is okay for me.'' A financial services lead consultant similarly described generating, scrutinizing, and revising: after prompting they ``scrutinises response, edit and put my input.'' These accounts frame prompting as part of an editorial workflow, where participants tighten constraints over turns, request specific formats, and treat each output as draft material to be tuned toward a locally acceptable result. When available, prompt histories participants shared during demonstrations illustrated how constraints were tightened over turns, which we treat as supportive context alongside interview accounts rather than a uniform observational record.

\subsubsection{Editing for originality, ownership, and voice as part of quality control}

Participants also described post-generation editing as a routine, especially to reduce recognizability and maintain ownership. A business development service provider emphasized deliberate rewriting: ``I make sure that I add my own diction, my own choice of words\ldots{} to the extent that it will be difficult, if not impossible for people to know.'' A power-sector trainer similarly described adding ``serious inputs'' to training manuals because of ``copyright, plagiarism and the rest,'' so they could ``claim it'' rather than ``picking it up as an AI generated information.'' In these accounts, editing is not cleanup. It is a strategy to make outputs socially and professionally viable.

\subsubsection{Boundary conditions: when steering is not enough}

Steering routines did not eliminate uncertainty about correctness, provenance, or traceability, especially for factual claims. Several participants described avoiding reliance on AI for particular kinds of facts. One participant stated: ``I don't use it mostly for facts\ldots{} for research I use Google,'' while another drew a clear boundary around numbers and market research: ``I don't rely on ChatGPT to get my statistics\ldots{} I rather go to Google and other industry websites to get exact statistics.'' Others described reverting to manual research when outputs felt untrustworthy: ``sometimes when I feel like I don't trust this, I have to go back and then do the manual method of research.''

A further friction was traceability. Participants reported that even when an answer sounded plausible, the stated sources could be unusable. A sales and marketing lead described clicking cited links and finding they ``always goes to 404,'' concluding: ``Your answer is not wrong, but your source\ldots{} the reference is so wrong, it's nonexistent.'' Another participant warned that when asked for references, ``some of the references\ldots{} may be fake'' and therefore ``you have to find out.'' These accounts show that interaction routines were not only techniques for improving output but also practices for deciding when AI-assisted drafting remained too risky or insufficiently accountable for the task at hand.

\subsection{Orchestrating an ecology of tools through switching and comparison}

Many participants described generative AI use not as stable commitment to a single platform, but as assembling an ecology of tools and moving between them, switching and comparing outputs as part of producing something usable for work. In these accounts, tool choice was not a one-time adoption decision but an interaction practice shaped by what participants believed each system could do well, where it failed, and how they managed uncertainty before acting on an output. Participants described this ecology through switching to exploit differentiated affordances (4.3.1), parallel prompting and triangulation across tools (4.3.2), reconciliation work when outputs diverged (4.3.3), and boundary conditions that triggered fallback to search/manual methods or paid tiers (4.3.4).

\subsubsection{Switching to exploit differentiated affordances}

Participants described switching tools when a platform did not meet the requirements of the task, especially around freshness and multimodality. A power sector trainer explained that preparing slide decks required ``diagrams,'' ``pictures,'' ``animations,'' and ``multimedia data,'' and that this was a reason they relied more on Bing when they needed richer visual material than they could readily get through a text-first workflow. Another participant framed switching as a response to currency, explaining that when they needed ``current information,'' they avoided relying on ChatGPT because (at the time, using ChatGPT 3.5) it was ``outdated'' and described its data as ending around ``September 2021''; for more recent information, they turned to Bing Chat or Google. Across these accounts, switching functioned as situational matching, routing work to the system participants believed best fit the deliverable constraints and immediate requirements of the task.

\subsubsection{Parallel prompting and triangulation}

Some participants described running the same prompt across multiple tools and comparing outputs. A consultant in financial services explained: ``If this tool is giving me wrong, I go to another tool\ldots{} I'm using three ai tools, and I give all the three tools the same prompt.'' In these accounts, cross-tool comparison was described as part of the work, and agreement or divergence across systems served as information for what to trust, revise, or verify further.

Others described triangulation as a way to consolidate overlap and then adapt differences for a specific audience. A power sector trainer described prompting multiple ai systems, adopting what appeared common across them, and then shaping differences through their own judgment of what would fit their trainees.

\subsubsection{Arbitration when outputs diverge}

Multi-tool comparison did not remove the need for human judgment. When outputs differed, participants described doing reconciliation work to decide what fit the task and the audience. A power sector trainer described this arbitration step directly: ``where there are differences or similarities, I now coined it to what I think will fit better for my trainers.'' Rather than resolving uncertainty automatically, cross-tool use introduced an additional synthesis step where participants had to choose, adapt, and rewrite before a result became usable.

A related boundary condition is that plausible alternatives can proliferate, leaving users to decide what to stand behind. One legal professional described challenging an answer and receiving ``a different version,'' then asking which story should be believed.

\subsubsection{Boundaries, fallbacks, and paid tiers}

Participants also described limits to multi-tool orchestration and named situations where switching did not resolve accountability demands. Several participants drew clear boundaries around factual queries and research, where they preferred external sources with traceable grounding. A CEO in the creative writing industry explained: ``I don't use it mostly for facts\ldots{} I won't ask it, for example, who is the president of Ghana\ldots{} I don't use it for research. For research I use Google.'' They added that while they had not personally experienced errors, they had ``heard of cases'' where the system gives a wrong answer and confidently explains it. Here, the boundary is an explicit rule of use that routes factual and research work back to search rather than cross-tool comparison.

Cost and access constraints also shaped when participants stayed within the tool ecology versus escalating to paid tiers. An IT company cofounder framed payment as a time tradeoff: ``if it's gonna save me eighty hours, why not?'' A sales and marketing lead described willingness to pay in a task-scoped way, tying payment to a specific deliverable type (presentations) and a bounded usage period. A telecom vendor similarly discussed ``limitations'' of free access and positioned upgrading as a way to gain speed and additional features they considered useful as the company's needs grew.

Across these boundary cases, participants treated switching as one strategy among others, bounded by accountability demands and by the costs, limits, and access conditions that shape whether multi-tool comparison is viable in everyday work.

\subsection{Accountability practices: verification and making outputs safe to ship}

Participants described that AI outputs were rarely treated as ready-to-use artifacts when correctness, traceability, or reputational risk mattered. Rather than framing ``trust'' as a stable attitude, they described concrete accountability practices that determined whether an output could be put into workplace use. We treat these accountability practices as release work: routines that make outputs safe to ship. In this theme, we describe how participants identified when verification was necessary (4.4.1), carried out cross-checking routines before use (4.4.2), encountered traceability breakdowns when references failed (4.4.3), and set non-delegation boundaries for high-stakes decisions, routing work back to primary sources (4.4.4).

\subsubsection{Verification triggers}

Participants described verification as necessary because AI output could be dated, incomplete, or misaligned with domain requirements, especially when the work had external accountability. A legal services partner drew a sharp boundary for legal research, arguing that the tool is ``dated'' and that they would not rely on it ``until and unless I see a statement from AI or ChatGPT saying we undertake to check with parliamentary reports and update our information.'' In their practice, what triggers verification is not only fear of being wrong, but the absence of an auditable commitment to keeping core sources current.

Other participants framed the trigger as the need to defend claims. A sales and marketing lead explained that even when the output seems ``quite correct,'' they still ``have to go and do research to get the reference points to defend what it has outputted.'' In these accounts, verification is activated by external accountability, where claims must hold up to likely challenge, not just sound plausible.

Participants framed verification as a gate triggered by domain standards and external accountability, not only by generalized mistrust.

\subsubsection{Cross-checking routines}

Participants described verification as a routine they perform before using or sharing AI-assisted work. For some, this meant routing factual and research work back to search. A CEO in the creative writing industry explained: ``I don't use it mostly for facts\ldots{} I don't use it for research\ldots{} for research I use Google.'' Here, ``verification'' is operationalized as tool switching, where AI remains a productivity aid while external sources carry the burden of factual grounding.

Other routines were more procedural and collective. The legal services partner described an established internal practice where a librarian checks authoritative sources ``with the Gazette and with the government printers to ensure Parliament hasn't updated or has updated'' legislation, and instructed their team to treat AI only as a starting point and then ``go back to your\ldots{} primary resources, your legislation, your statutes, your government reports, your parliamentary\ldots{}'' This makes verification visible as coordinated work that uses institutional roles and primary-source checking, rather than individual caution alone. In practice, verification operated as release work, combining tool-switching, primary-source checking, and, in some settings, distributed review roles.

\subsubsection{Traceability breakdowns}

Several participants described a specific accountability problem: even when an answer appeared reasonable, it did not reliably come with retrievable sources. Traceability failures became most salient when outputs were intended for external circulation, where participants anticipated being asked to justify claims rather than using the content for internal exploration. A sales and marketing lead articulated this as a need for post-hoc traceability work, explaining that they still ``have to go and do research to get the reference points to defend what it has outputted.'' In demonstrations where available, participants pointed to generated citations or links in outputs as illustrative cases of sources they could not access, which we treat as supportive context alongside interview accounts rather than a uniform observational record. These accounts describe a traceability gap: tools could generate plausible claims but did not reliably supply the evidence trails participants needed for accountable workplace use.

In this sense, verification is not only checking for correctness. It is reconstructing a checkable justification, especially when the output will be used in professional settings where others may demand ``where did this come from?''

\subsubsection{Non-delegation boundaries}

Participants also described boundaries around what they would not delegate to AI, particularly when consequences were high-stakes or when the domain required authoritative grounding. A research and development lead in printing and design emphasized that they would ``never allow it to make financial decisions'' without them, describing AI as potentially useful for analysis or surfacing missed correlations, while insisting that ``the financial decisions\ldots{} still remain with me.'' This frames accountability as non-transferable, even when AI contributes analytical input.

A CEO in an outdoor recreation company similarly described an organizational gatekeeping pattern: AI can generate drafts, but ``someone has to check factual,'' and outputs cannot be copied directly into work ``proform[a]'' without review. Together with the legal partner's refusal to rely on AI for up-to-date law absent verifiable updating practices, these accounts show that participants draw boundaries where responsibility, consequences, or source authority demand human control.

These boundaries clarify the division of labour: AI can accelerate drafting and analysis, but responsibility for high-stakes judgment and factual accountability remains with the worker.

\subsection{Context misfit and repair: localization work when outputs do not travel}

Participants described breakdowns that were not about generic ``accuracy'' alone, but about whether AI outputs fit local audiences, languages, and domain realities. Misfit showed up when outputs sounded unlike the speaker, when tools mishandled local words and accents, when responses defaulted to Western or Americanised assumptions, and when generated media failed to represent African people and places convincingly. Participants responded through repair work: rewriting for voice, tightening prompts with local constraints, substituting local sources, switching tools, and, in some cases, abandoning a capability entirely when localization failed.

\subsubsection{Sounding like me}

Even when outputs were fluent, participants often treated them as drafts that needed reshaping to match personal voice and audience expectations. A sales and marketing lead described a common workflow: ``i just type in what i want to say or my key points, then it dashes out the data... then i edit it to sound like me. and be presentable like me.'' Others revised because the output felt overly formal or unnatural. An operations manager noted that responses can be ``too technical\ldots{} you seem like you're a robot,'' requiring ``human elements,'' and explained how they would still revise: ``there's no human element\ldots{} I would, still take it and revise it again\ldots{} to how [I] would, say this?''

For these participants, ``fit'' was not just stylistic polish. It was part of making the message socially safe for the intended audience, especially when the writing would represent them professionally.

\subsubsection{Language breakdowns}

Localization breakdowns became most visible when tools encountered local languages, names, and accents. A CEO in outdoor recreation explained they ``had to abandon it'' because their work increasingly involved ``a lot of local languages,'' and the system struggled with pronunciation and naturalness. They described how it could handle global place names better than local ones: ``it's not able to pronounce it as well as\ldots{} Nairobi\ldots{} London\ldots{} Everest\ldots{} when it comes to local names, it's not able to localise them\ldots{} verbalise them.''

A product designer described similar issues in text-based prompting and transcription, noting that ``Sometimes it doesn't really understand those African words\ldots{} African accents really well,'' and that misrecognition could shift meaning: ``it reads one word wrong and that changes the whole meaning.'' Some participants described adjusting communication practices when using these tools, for example avoiding code-mixing or switching to English.

When language handling fails, repair becomes structural rather than cosmetic, shaping whether the tool can be used at all.

\subsubsection{Locality gaps}

Misfit also appeared when outputs failed to reflect local domain realities, even when systems were connected to online sources. A head of business development in legal services described probing for culturally specific knowledge and encountering unstable answers. When they asked for the Meru people's ``Genesis version,'' they reported that the tool returned one version, then after correction produced another, then another: ``so which one do you believe?\ldots{} story A\ldots{} or story B\ldots{} and another version appeared\ldots{} it is not a very good research tool.''

In legal work, locality carried high stakes. A partner in a legal service argued that even internet-connected tools were ``still unable to [get] Kenya\ldots{} the information,'' producing outputs that were ``really Americanised,'' and missing contextually decisive local materials such as a scanned case that changes statutory interpretation: ``it cannot make that link\ldots{} even though it's connected to the Internet, it's still churning out incorrect information.'' A related incident came from an associate who described a team exercise where AI-generated quoted law contained ``a lot of red flags,'' concluding it ``destroy[ed] my confidence in the platform.''

Others treated locality gaps as a reason to use outputs only as scaffolding. A research and development lead in a printing firm described generating ``general text'' and then tailoring it: ``we personalized it to our field\ldots{} but at least we got the generic part\ldots{} Again, I had to help them with this.''

Local fit, in these accounts, is produced by substituting or layering situated knowledge onto generic drafts, and by rejecting outputs outright when local specificity is non-negotiable.

\subsubsection{Representation and repair}

Misfit also surfaced in representational outputs, especially images and synthetic voices. A CEO in outdoor recreation described prompting text-to-image with ``a black man sliding in a hike'' and finding the result visually incoherent: ``you can't really say this is real black\ldots{} a black face and white man's hair,'' and noted that a ``landscape photo in Kenya'' could yield non-African scenery, ``showing snow.'' In cases where participants shared example image outputs, they used these artifacts to illustrate representational incoherence and locality mismatch. A CEO and founder in ICT services described a similar representational failure in speech, noting that even voice cloning produced ``a British accent or\ldots{} an American accent,'' adding: ``I'm not British,'' and arguing that ``we need more African voices and real African voices.''

Repair was often described as constraint layering and iterative regeneration rather than one-shot prompting. The outdoor recreation CEO explained that being more specific helped, for example by naming the exact destination: ``an itinerary for hikers going to Mount Kinango\ldots{} So not leave it open.'' Others framed the path to acceptable outputs as repeated experimentation: ``keep experimenting and keep tweaking the prompts, then eventually you get what you want.'' In some cases, repair was not worth it. The outdoor recreation CEO described paying for a month and then stopping: localization failures made it ``not\ldots{} sounding original,'' and training the system felt ``quite a long process,'' leading them to conclude the tool ``wasn't really ready\ldots{} for our market.''

Context misfit, in the form of localization and representational failures, surfaced as concrete breakdowns in credibility, and repair took the form of constraint layering, regeneration, tool switching, or disengagement when misfit persisted and outputs could not be made defensible for circulation.

\section{Discussion}

\subsection{Drafting and release work}

Across our accounts, participants treated generative AI as draft labour that accelerated first-pass production of materials intended for circulation, followed by revision and verification. In these accounts, `circulation' often meant client-facing release where errors carry reputational or commercial risk. Yet, participants described perceived speed gains as conditional on whether they could make AI-assisted drafts safe to ship for client-facing circulation through release work. Our accounts foreground the accountability costs that controlled task-completion measures can miss, extending prior task-level evidence on performance improvements and error modes \cite{brynjolfsson2023,dellacqua2023,noy2023}.

We define release work as boundary-facing accountability labour that gates client-facing circulation under external accountability. In our data it appears in three empirical markers: verification gates tied to domain standards, post-hoc traceability reconstruction when sources fail, and non-delegation boundaries that allocate responsibility for what ships. Release work is closely related to articulation work, and we do not treat it as lying wholly outside that tradition. Articulation work has long been understood to extend beyond scheduling interdependent tasks to the integration, alignment, and quality-assurance labour that makes cooperative work cohere \cite{schmidt1992}. Release work is not reducible to it, for three reasons. Its object is narrower: not the mesh of many tasks and actors, but the defensibility of a single artifact about to cross an organizational boundary. Its trigger is different: not task interdependence among contributors, but external accountability at the moment of circulation. And, decisively, it can be discharged by one worker with no cooperative activity to articulate.

Our own data show both cases side by side. Where the legal partner routed verification through a librarian who checks legislation against the Gazette and instructed staff to return to primary sources (Section 4.4.2), verification was accomplished as coordinated, distributed work, that is, as articulation work. But a worker who alone certifies that a draft is defensible enough to send performs release work with no interdependent contributors to coordinate, as when a financial services consultant cross-checked every response "before I use it" (Section 4.1.4) and an R\&D lead held that "the financial decisions still remain with me" (Section 4.4.4). With owners and founders comprising twelve of our twenty-seven participants and client-facing roles twenty-one, this single-worker certification is common rather than incidental in our sample. The distinction matters because participants described effort shifting from producing text to producing accountability as outputs headed toward circulation, especially when tools generated plausible claims without usable evidence trails.

In contrast to task-completion framings that treat ``quality'' as an output property assessed at the end of a task, our accounts foreground the extra work required to make an AI-assisted draft safe to ship for circulation. The key comparison is not only whether generation shortens drafting time, but whether workers can assemble the evidentiary scaffolding and accountability boundaries that let a draft travel into circulation without creating credibility or liability exposure. On this view, participants' perceived speed gains were conditional on release work being feasible within deadlines and institutional standards, and on workers being able to reconstruct traceability when tools did not surface retrievable sources.

In our accounts, release work often took the form of routing claims back to traceable sources and setting explicit boundaries for what could not be delegated. The next sections show how steering and tool ecologies supported release by making drafts easier to verify.

\subsection{Steering as editorial practice}

Participants rarely described getting usable value from a one-shot prompt. Instead, they developed interaction routines to produce ``quality'' through meta-prompting, scaffolding with templates or add-ons, iterative refinement loops, and post-editing. This complements workplace studies suggesting that generative AI is often used as a support tool and that workers anticipate uneven transformation across domains \cite{sun2024,wagman2025,woodruff2024}. We specify how this support is accomplished in day-to-day work. Participants treated prompting as an editorial workflow: they tightened constraints over multiple turns and then rewrote outputs to meet client-facing standards. These routines helped participants produce drafts they were willing to release.

A notable feature of these routines is that ``quality'' includes ownership and voice. Participants described rewriting to reduce recognizability and to preserve their diction, both to maintain credibility and to manage plagiarism concerns. In our data, this was not only stylistic preference. It is part of making outputs circulable, particularly where tone and authenticity are professionally consequential. This reframes ``prompt skill'' as a situated quality-production practice embedded in accountability and credibility concerns, rather than a transferable individual competence \cite{orlikowski2000,suchman1987}.

At the same time, steering did not resolve all uncertainty. Participants reported boundaries around facts, numbers, and research, and described traceability failures, including unusable citations, that reintroduced manual research as a fallback. These limits reinforce that ``better prompting'' does not remove the need for accountable grounding. Instead, it redistributes where effort is spent and how responsibility is managed in everyday production. In practice, these steering routines supported release by shaping drafts into forms participants could credibly stand behind before circulation.

\subsection{Tool ecologies and arbitration}

Participants did not describe generative AI use as stable commitment to a single platform. They assembled ecologies of tools and moved between them to exploit differentiated affordances, pursue freshness, obtain multimodal assets, and triangulate outputs across systems. CSCW has long described infrastructures as heterogeneous, where workability is achieved by assembling and navigating multiple resources rather than relying on any single system's completeness \cite{monteiro2013,star1996}.

In our accounts, cross-tool comparison functioned as a routine for managing uncertainty. Some participants ran the same prompts across multiple systems and treated agreement, divergence, and overlap as signals for what to trust, revise, or verify further. However, comparison did not eliminate judgment work. Divergence introduced arbitration, where participants had to reconcile alternatives into one deliverable that fit the task and the audience and that they were willing to stand behind. This arbitration is part of the operating cost of ecological use, and it is visible in our Findings as reconciliation work that follows triangulation rather than replacing it.

Evaluating a single tool in isolation can miss how work was routed across tools, search, and paid tiers to manage uncertainty and meet deliverable constraints. Participants treated this workflow ecology as a quality-control and release resource: comparison and search fallback enabled progress, but divergence introduced arbitration overhead.

\subsection{Context misfit and repair}

Across our interviews, breakdowns were not only about factual correctness. Participants described context misfit when outputs failed to match local requirements, including personal voice, local languages and accents, local domain realities, and credible representation of African people and places. In circulation, these misalignments mattered because they created credibility risks, could distort meaning, and generated additional repair work before participants felt comfortable using the output externally. We operationalize context misfit in participants' accounts when they describe localized breakdowns that demand repair to meet workplace standards, or when they abandon a capability because repair costs outweigh value. These dimensions are not intended as exhaustive. They reflect the recurring misfit patterns across participants' accounts.

These dynamics connect to prior work that frames related issues as geographic and cultural representational bias and long-tail knowledge gaps in generative systems \cite{basu2023,kandpal2023,manvi2024}, as well as community-centered accounts showing how text-to-image systems can reproduce narrow representational regimes \cite{alenichev2023,qadri2023,zhang2024}. Our findings extend this literature by showing how these failures surface in day-to-day work as breakdowns that require repair and, at times, make use not worth the effort. In our accounts, these failures became consequential through repair demands, credibility risks, and capability abandonment in situated workplace use.

A longer line of CSCW and geographic HCI research helps explain why these locality failures are not only a property of any single model but reflect a well-documented pattern in how digital geographic knowledge is produced. Two strands are relevant. First, the content of the digital record is geographically uneven: much geotagged user-generated content about a place is produced by non-locals rather than residents, so it can encode an outsider vantage rather than situated local knowledge \cite{hecht2010}; peer-produced knowledge systematically underrepresents rural areas in both coverage and local authorship, leaning more on automated agents where local contributors are scarce \cite{johnson2016}; and crowd-sourced map coverage tracks population density and socio-economic status rather than need \cite{mashhadi2013}. Second, the participation that generates this record is itself unequal: the crowd-labour markets used to collect situated data can avoid lower-income areas \cite{thebaultspieker2015}, and even crisis-driven mapping shows steep participation inequality \cite{dittus2017}. We do not claim that these specific datasets are the training corpora of the tools our participants used. Rather, generative systems are trained on web-scale content shaped by the same production and participation dynamics, and they inherit the same structural unevenness. Our contribution is to trace this inheritance to the point of use: we show how the resulting gaps resurface in everyday SMB work as concrete breakdowns that demand repair, credibility management, and, at times, capability abandonment before outputs can be released.

Language and speech misfit also show that localization challenges span modalities. Prior work documents systematic disparities in speech technologies across race, dialect, and language variety, including in automated captions and recognition performance \cite{koenecke2020,tatman2017}. In our interviews, participants described analogous frictions in generative voice tools: accent handling and local-name pronunciation shaped whether outputs were usable in client-facing settings. Participants reported mispronunciations and accent misrecognition that could change meaning, and in some cases led teams to avoid code-mixing or default to English. One participant described voice cloning that produced a ``British'' or ``American'' accent rather than a locally recognizable voice, motivating their call for ``real African voices.'' These accounts suggest that localization is not only a translation problem but also a question of whether speech interfaces can produce locally credible vocal identity and pronunciation in everyday work (Sections 4.5.2 and 4.5.4).

Participants also encountered locality gaps in institutional and cultural knowledge, including legal contexts where locally authoritative materials and interpretive context were decisive. In Section 4.5.3, legal participants described ``Americanised'' responses and missing links to locally decisive materials (e.g., scanned cases), which in these accounts made the tools difficult to rely on for research tasks where local specificity is required. Here, misfit was not only a matter of tone or lexical choice. Participants described outputs that defaulted to non-local assumptions and were difficult to ground in local sources, increasing verification and substitution work. This aligns with calls to evaluate effectiveness in culturally nuanced, real-world scenarios rather than relying only on decontextualized benchmarks \cite{adilazuarda2024,ochieng2024}.

Participants responded to context misfit through repair: rewriting for voice, constraint layering, substituting local sources and situated knowledge, switching tools, and in some cases abandoning capabilities when localization failures persisted. Sections 4.5.2--4.5.4 illustrate how these repairs played out in practice, including disengagement when repair costs outweighed value. This makes breakdown and repair visible as ongoing integration work that sits between generation and circulation \cite{schmidt1992,suchman1987}. Kenya and Nigeria were especially revealing here because participants' accounts repeatedly surfaced linguistic, institutional, and representational requirements that made repair work especially visible between generation and client-facing circulation.

\subsection{Workable participation}

Participants who were not AI specialists used generative tools to draft, refine, and package workplace materials, including in ways that supported professional communication and participation. Arguments that democratization has multiple meanings, and that access alone does not guarantee equitable benefit, bear directly on this \cite{capraro2024,seger2023}. Scholarship on emerging AI divides makes the same point, emphasizing the uneven conditions through which people convert access into advantage \cite{bentley2024,daepp2024,vassilakopoulou2023}.

In this paper, we operationalize democratization as workable participation: the ability to produce locally credible, accountable outputs with acceptable time, verification effort, and risk. Across our accounts, workable participation was evidenced when participants could complete the release-facing accountability work described in 5.1 within delivery timelines, including verification, traceability reconstruction for key claims, and clear non-delegation boundaries for residual risk. These diagnostics surface across our findings, including the release-work episodes in Sections 4.4.1--4.4.3 and the localization and repair cases in Sections 4.5.2--4.5.4. This sharpens ``democratization'' beyond access or adoption by focusing on whether workers can make outputs ready for accountable circulation, rather than merely generate drafts. On this definition, access can expand drafting, while the distribution and cost of release work and localization work shaped who could reliably benefit under these accountability conditions \cite{bentley2024,daepp2024,seger2023}. In consequential work, participants described needing retrievable sources, traceable grounding, and organizational gates before outputs could be released. When traceability failed, workers described reconstructing evidence trails after generation, including cases where cited sources were unusable. In local-facing work, participants described misfit across language, locality, voice, and representation that required repair and, at times, led to capability abandonment. We treat workable participation as a situated analytic lens for these interview settings, not a general metric of democratization across workplaces.

Building on the misfit cases in 5.4, workable participation often required substitution and layering: routing claims back to primary sources, adding situated knowledge to generic drafts, tightening prompts with local constraints, switching tools, or disengaging when repair costs outweighed value. Put differently, access expanded the possibility of drafting, but workable participation depended on whether workers could stabilize verification and localization within acceptable time, verification effort, and risk.

\subsection{Contributions}

Taken together, our findings make two central contributions to CSCW research on generative AI at work, connected by a reframing of democratization and supported by a five-practice account and design directions. As a foundation, we have shown how participants made generative AI workable through five intertwined practices: draft labour for deliverables headed to external circulation, steering quality through interaction routines, orchestrating tool ecologies through switching and comparison, performing release work through verification, traceability reconstruction, and non-delegation boundaries, and repairing context misfit when localization and representational failures had to be resolved before outputs could circulate. Drafting was routinely paired with steering, cross-tool comparison, and release work as drafts moved toward circulation.

Our first contribution is conceptual. We specify release work as a distinct, boundary-facing form of accountability labour that determines whether an AI-assisted draft can circulate to an external audience, evidenced through verification gates, traceability reconstruction when references fail, and non-delegation boundaries that allocate responsibility for what ships. As we argue in Section 5.1, release work is related to articulation work but not reducible to it: it takes a single circulating artifact as its object, is triggered by external accountability at the point of circulation rather than by task interdependence, and can be performed by one worker with no cooperative activity to coordinate. This clarifies why perceived speed gains were reported as contingent on whether release work could be completed within deadlines and institutional standards.

Our second contribution is empirical. We characterize context misfit as a recurring breakdown-and-repair class that gates circulation, localization and representational failures that become consequential when outputs cannot be made locally credible for external circulation without substantial repair. This is a mechanistic contribution, not a taxonomic one. We do not claim a new category of bias, since prior work already documents geographic, cultural, linguistic, and representational bias. What we add is an account of how those failures operate in everyday work, where prior accounts often treat them as properties of outputs, whereas in our data they became consequential through the repair work and abandonment decisions they prompted. We document a repair repertoire that includes rewriting for voice, layering constraints, substituting local sources and situated knowledge, switching tools, and disengaging when repair costs outweighed value.

We connect these contributions through workable participation, a reframing of democratization from access toward the ability to produce locally credible, accountable outputs within acceptable time, verification effort, and risk. We translate them into four design directions that target release work, steering, ecological use, and localization.

\subsection{Design implications}

A single principle runs through these directions: design for the moment an output leaves the worker and crosses to a client, not for the moment it is generated. In our accounts consequential friction appeared at circulation, so the design target is the boundary crossing rather than first-draft production.

Make provenance legible before circulation, not reconstructed after it. The problem participants described was not citation accuracy but that fluent claims arrived with no defensible, checkable trail, forcing post-hoc reconstruction under external accountability (Sections 4.4.1 to 4.4.4, especially 4.4.3). Two moves follow. First, a provenance-first output mode that marks each factual assertion as either grounded in a retrievable source or explicitly unverified, so the verification gate is applied up front rather than rebuilt later. Second, a portable evidence trail generated alongside the draft, including source, retrieval date, and a fallback query, that the worker can attach when the artifact crosses to a client, turning private checking into a shareable accountability record. The failure mode to avoid is the naive one: adding citation features that manufacture confidence and reproduce the dead or fabricated references participants already distrusted, links that "always goes to 404." The goal is to make the absence of grounding visible, not to smooth it over. 

Treat voice as an owned, cumulative asset, not a per-session setting. Participants rewrote outputs to manage authorship, credibility, and plagiarism risk, not merely to adjust tone, and treated learning how to ask as work in itself (Sections 4.1.3, 4.2.1, 4.2.3 to 4.2.5). One option is a user-owned, inspectable voice profile that the worker can edit and carry across sessions, so steering compounds rather than restarting each time. Because prompt formulation was itself learned labour, a prompt-formulation scaffold that turns a vague goal into a structured request would address that elicitation labour directly. The tension is homogenisation: such a profile must remain inspectable and easy to override so it does not over-fit past text \cite{sun2024,woodruff2024}.

Support arbitration across tools, not only comparison. The costly work was not noticing that outputs differed but deciding which to trust and reconciling them into one audience-specific deliverable (Sections 4.3.2 to 4.3.4, especially 4.3.3). Rather than difference-highlighting alone, systems could explore ways to surface agreement and divergence at the level of specific claims, so divergence pinpoints what to check. The caveat that must be designed in: because similar tools can share the same blind spots, agreement should be treated as a heuristic for where to look, not as verification \cite{monteiro2013,schmidt1996,star1996}. 

Design for local grounding and for honest early failure. The deepest misfits were institutional and place knowledge and representation, not translation, and they often ended in paid-for abandonment (Sections 4.5.2 to 4.5.4). Because workers already repair by injecting local materials, systems could let them ground generation in their own sources, a scanned case, a local corpus, or place-specific references, and constrain outputs to them, formalising the substitution and layering participants did by hand. Locale, voice, and representation constraints should persist and be inspectable. Critically, when a model cannot meet a locale or representation requirement it should say so up front, so workers can disengage before sinking repair cost and subscription money rather than discovering the limit after a month, as one participant did. The tension is tokenism: representation controls should draw on community-grounded defaults rather than a set of surface sliders \cite{koenecke2020,manvi2024,ochieng2024,qadri2023}.

\subsection{Scope and limitations}

This study examines reported workplace practices among SMB workers in Kenya and Nigeria and uses these sites for analytic leverage rather than representativeness. Because recruitment relied on professional networks and required recent workplace GenAI use, the sample reflects digitally connected workers already using these tools and should not be read as prevalence or adoption estimates. Interviews were conducted in English, which may privilege participants with higher comfort using English at work and may under-observe prompting and collaboration practices that occur primarily in local languages. We do not make claims about differences between the two sites. Because fieldwork took place between July and October 2023, the findings are a snapshot of early workplace use during a rapidly changing period. Tool capabilities, interfaces, pricing, and failure modes have since evolved, so the study should not be read as an assessment of current model performance. Finally, our data does not support causal claims about model development or tuning (e.g., reinforcement learning from human feedback, RLHF) or about training data composition as direct explanations for the breakdowns we report. We discuss these mechanisms only as plausible contributing factors rather than findings from our data. Future work can test how these practices vary across language regimes, governance requirements, and tool configurations that shape traceability and localization performance.

\section{Conclusion}

This study examined how SMB workers in Kenya and Nigeria incorporate generative AI into everyday work, and how they manage reputational and commercial risks when AI-assisted drafts must be made safe to ship in client-facing work. Across accounts, generative AI became workable through five interlocking practices: using AI as draft labour for client-facing deliverables, steering quality through interaction routines and post-editing, switching across tool ecologies through comparison and arbitration, performing accountability practices for verification, traceability, and non-delegation, and repairing context misfit when local language practices, institutions, or representation do not align with workplace standards.

These findings answer our research questions. For RQ1, incorporation was an ongoing practice organized around verification triggers, accountability boundaries, and cross-tool comparison that shaped what could be safely released in client-facing work. For RQ2, we characterize context misfit as a recurring breakdown class and document a repair repertoire that includes rewriting for voice, localization work, substituting situated knowledge, switching tools, and abandoning outputs when repair costs exceed value.

For CSCW, our findings show that reported gains were conditional on accountability routines, particularly the release work required to make outputs defensible for circulation. In these accounts, the central analytic object is the verification, arbitration, and repair infrastructure that shaped everyday use. Designing workplace AI that reduces localization and accountability burdens in such settings motivates four design directions: support for verification and traceability, iterative steering and post-editing, ecological comparison and arbitration across tools, and localization beyond language, including representational fidelity. These claims describe practices in our two sites rather than representative global patterns.

\bibliographystyle{ACM-Reference-Format}
\bibliography{references}

\end{document}